\documentclass[journal]{IEEEtran}

\usepackage{amsmath}
\usepackage{url}
\usepackage{graphicx}
\usepackage{color, soul}

\begin{document}

\title{Assessing the Accuracy of a Wrist Motion Tracking Method
for Counting Bites across Demographic and Food Variables}

\author{Yiru Shen,
        James Salley,
        Eric Muth,
        and Adam Hoover~\IEEEmembership{Senior Member, IEEE}
\thanks{Y. Shen and A. Hoover are with the Department of Electrical
and Computer Engineering, Clemson University, Clemson, SC 29634-0915 USA
(e-mail:  yirus@clemson.edu; ahoover@clemson.edu).}
\thanks{J. Salley and E. Muth are with the Department of Psychology,
Clemson University, Clemson, SC 29634 USA
(e-mail: jnsalle@gmail.com; muth@clemson.edu).}
}


\maketitle

\begin{abstract}
This paper describes a study to test the accuracy of a method
that tracks wrist motion during eating to detect and count bites.
The purpose was to assess its accuracy across demographic (age, gender,
ethnicity) and bite (utensil, container, hand used, food type) variables.
Data were collected in a cafeteria under normal eating conditions.
A total of 271 participants ate a single meal while
wearing a watch-like device to track their wrist motion.
Video was simultaneously recorded of each participant and subsequently
reviewed to determine the ground truth times of bites.
Bite times were operationally defined as the moment when food or beverage
was placed into the mouth.
Food and beverage choices were not scripted or restricted.
Participants were seated in groups of 2-4 and were encouraged to eat
naturally.
A total of 24,088 bites of 374 different food and beverage items were
consumed.
Overall the method for automatically detecting bites had a
sensitivity of 75\% with a positive predictive value of 89\%.
A range of 62-86\% sensitivity was found across demographic variables,
with slower eating rates trending towards higher sensitivity.
Variations in sensitivity due to food type showed a modest correlation
with the total wrist motion during the bite, possibly due to an increase
in head-towards-plate motion and decrease in hand-towards-mouth motion
for some food types.
Overall, the findings provide the largest evidence to date that the method
produces a reliable automated measure of intake during unrestricted eating.
\end{abstract}


\section{Introduction}

\IEEEPARstart{M}{ore} 
than half of the world population is overweight (39\%) or
obese (13\%) \cite{WHO2016}.
Obesity is associated with increased risks for cardiovascular disease,
diabetes, and certain forms of cancer \cite{Malnick06}, and has become
a leading preventable cause of death \cite{Mokdad04}.
The study and treatment of obesity is aided by tools that measure
energy intake, determined by the amount and types of
food and beverage consumed.
Existing tools include questionnaires about the frequency of food
consumption, food diaries, and 24-hour recalls of the foods
consumed during the day \cite{Day01,Thompson08}.
However, these tools rely upon self-report and have a number of
limitations, including high user and experimenter burden, interference
with natural eating habits, decreased compliance over time, and
underreporting bias \cite{NIH2012,Thompson08}.
Experts in the field of dietetics have emphasized the need for
technology to advance the tools used for energy intake
monitoring \cite{McCabe-Sellers10,Schoeller13,Thompson10}.

Advances in body sensing and mobile
health technology have created
new opportunities for empowering people to take a more active role
in managing their health \cite{Kumar13}.
Wearable sensors have significantly advanced
the assessment of energy expenditure in the form of
accelerometer-based physical activity monitors \cite{Westerterp09}.
However, the development of a similar tool for monitoring
energy intake has remained elusive.
Researchers have investigated the automatic recognition of foods
in images \cite{Anthimopoulos15, He15,Pouladzadeh14,Zhu15}
and sensors worn on the throat and ear
area to detect swallowing events \cite{Amft09,Passler11,Passler14,
Sazonov08,Sazonov10b,Sazonov12}.
Our group has been investigating using a wrist-worn configuration of
sensors to detect periods of eating \cite{Dong14} and track
hand-to-mouth gestures \cite{Dong12,Ramos15}.
One benefit of wrist-mounted sensors is that they can be embodied in
a device that resembles a common watch.
This makes the monitoring inconspicuous which helps promote
long-term daily use \cite{Coons12}.

In previous work our group developed a method that detects a pattern
of wrist motion during the ingestion of a bite \cite{Dong09,Dong12}.
An experimental evaluation of 49 people eating a meal of their choice
in a laboratory setting found that the method counted bites with a
sensitivity (ratio of true detections to total actual bites) of 86\%
and a positive predictive value
(ratio of true detections to true detections plus false positives)
of 81\% \cite{Dong12}.
The experiment also revealed that an inexpensive micro-electro-mechanical
systems (MEMS) gyroscope was as accurate as a more sophisticated
magnetic, angular rate and gravity (MARG) sensor in tracking the
relevant motion pattern \cite{Dong12}.
These experiments were conducted using wrist-worn devices that were
tethered to a stationary computer in order to facilitate the recording
of raw motion data.
Subsequently, the method was instantiated in a wearable version that
resembles a watch.
The watch executes the algorithm to detect the relevant motion pattern
on a microcontroller.  A button is pressed at the beginning of an
eating activity (e.g. meal or snack)
to begin bite counting, and pressed again at the end of the eating
activity to end bite counting.
The total bite count for the eating activity is stored for subsequent
downloading to an external computer.
To test its relevance for measuring energy intake, 77 people wore the
device for 2 weeks and used it to
automatically count bites during all eating activities \cite{Scisco14}.
Participants completed the automated self-administered 24 hour recall
to measure kilocalories consumed \cite{Subar12}.
A total of 2,975 eating activities were evaluated, an average of 39
per participant.
A comparison of automated bite count to kilocalories found an average
per-individual correlation of 0.53, with 64 participants having a correlation
between 0.4 and 0.7 \cite{Scisco14}.
This range of correlation is similar to what has been found in evaluations
of energy expenditure measured by accelerometer-based devices
(pedometers, physical activity monitors) \cite{Westerterp07}.

This paper describes an experiment conducted to further
evaluate the accuracy of the automated bite counting method.
The goal was to record a large number of people eating a wide variety
of foods and beverages to evaluate its accuracy in terms of demographic
variables (gender, age, ethnicity) and bite variables (food type,
hand used, utensil, container).
One approach to such an experiment is to script activities and ask
each participant to complete the script.
For example, a participant could be asked to consume 5 bites of 20
different types of food in a controlled order.
This approach has been taken in some other studies of eating activities
(e.g. \cite{Amft06,Passler14,Sazonov12}).
Advantages to this approach include limiting the set of food types,
simplifying the ground truth identification of events due to
the use of a controlled script, and ensuring an
equal quantity of each event type through repetition.
However, this is unnatural in terms of food choices,
eating pace, food order, and overall behavior during normal eating.
Instead, we instrumented a cafeteria setting.
Participants were allowed to select their own foods and eat naturally.
This resulted in unequal distributions of bite variables 
which is offset by recording a large number of participants.
Section \ref{methods} describes the experimental conditions and
Section \ref{results} describes the variations in the accuracy of
the bite counting method due to demographic and bite variables.

\section{Methods}
\label{methods}

\subsection{Instrumentation}

\begin{figure}
\begin{center}
\begin{tabular}{c c}
\includegraphics[height=1.2in]{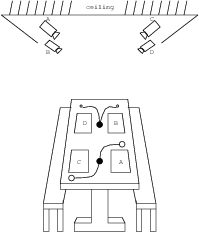} &
\includegraphics[height=1.2in]{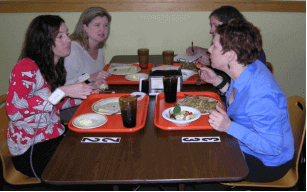} 
\end{tabular}
\end{center}
\caption{The table instrumented for data collection.
Each participant wore a custom tethered device to track wrist motion.}
\label{eating_table}
\end{figure}

The experiment took place in the Harcombe Dining Hall at Clemson University.
The cafeteria seats up to 800 people and serves a large variety of foods
and beverages from 10-15 different serving lines.
Figure \ref{eating_table} shows an illustration and picture
of our instrumented table \cite{Huang13}.
It is capable of recording data from up to four participants
simultaneously and is similar to others in the cafeteria so that
its appearance would not be distracting.
Four digital video cameras in the ceiling 
(approximately 5 meters height) were used to record each
participant's mouth, torso, and tray during meal consumption.
A custom wrist-worn device containing MEMS accelerometers
(STMicroelectronics LIS344ALH)
and gyroscopes (STMicroelectronics LPR410AL)
was used to record the wrist motion of each participant at 15 Hz.
Cameras and wrist motion trackers were wired to the same computers
and used timestamps for synchronization.
All the data were smoothed using a Gaussian-weighted window of
width 1 s and standard deviation of $\frac{2}{3}$ s.

\subsection{Participants}

The Clemson University Institutional Review Board approved data
collection and each subject provided informed consent.
A total of 276 participants were recruited and each consumed
a single meal \cite{Salley16}.
Participants were free to choose any available foods and beverages.
Upon sitting at the table to eat, an experimental assistant
placed the wrist motion tracking device on the dominant hand of the
participant and interviewed them to record the identities of foods selected.
The participant was then free to eat naturally.
If additional servings were desired, the participant was instructed
to notify the experimental assistant to assist with removing the
wrist motion tracker before moving through the cafeteria to obtain
more food or beverage, returning to the table to begin a new segment
of recording.
Each such segment is referred to as a course.
For 5 participants, either the video or wrist motion tracking data
failed to record, and so are excluded from analysis.
Total usable data includes 271 participants, 518 courses with a range of
1-4 and average of 1.8 courses per participant.
Demographics of the participants are 131 male, 140 female;
age 18-75; height 50-77 in (127-195 cm); weight 100-335 lb (45-152 kg);
self-identified ethnicity 26 African American, 29 Asian or Pacific Islander,
190 Caucasian, 11 Hispanic, 15 Other.

\subsection{Ground truth}

\begin{figure*}
\begin{center}
\includegraphics[width=0.9\textwidth]{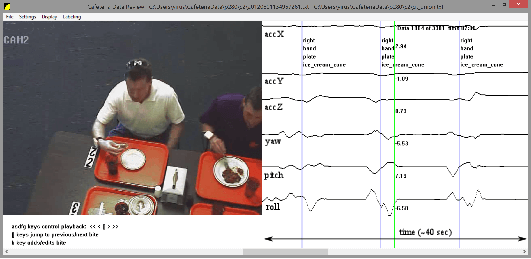}
\caption{A custom program created for manual labeling of ground truth bites.
The left panel shows the video and the right panel shows the wrist
motion tracking.  Vertical purple lines indicate the times marked as
bites, the vertical green line indicates the time currently displayed
in the video.  Variables (hand, utensil, container, food) are
identified for each bite.}
\label{tool}
\end{center}
\end{figure*}

\begin{figure*}
\begin{center}
\begin{tabular}{c c c c c}
\includegraphics[width=0.16\textwidth]{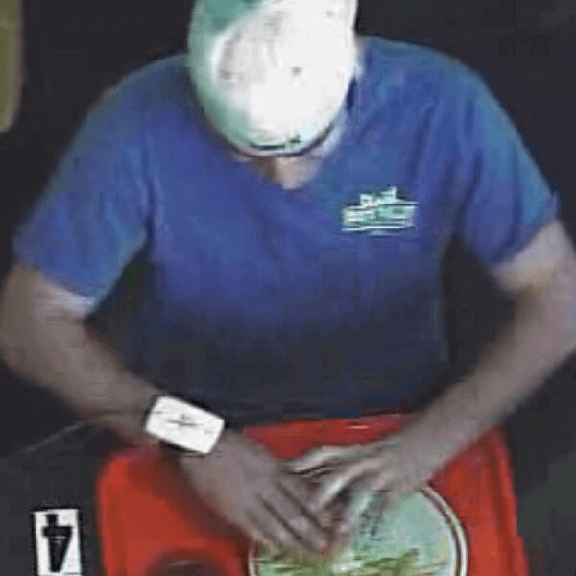} &
\includegraphics[width=0.16\textwidth]{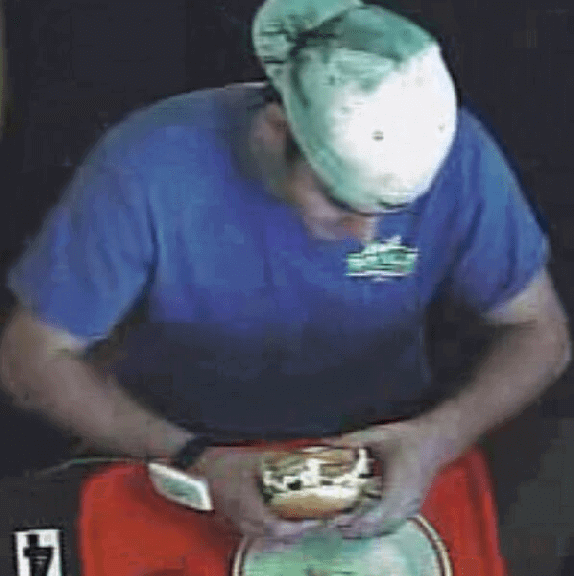} &
\includegraphics[width=0.16\textwidth]{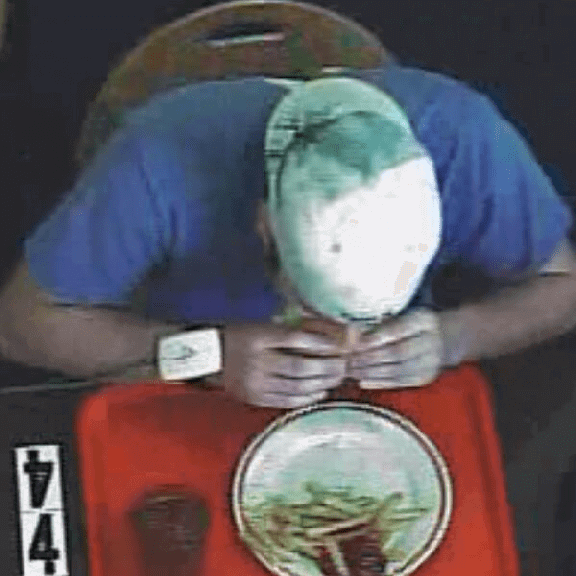} &
\includegraphics[width=0.16\textwidth]{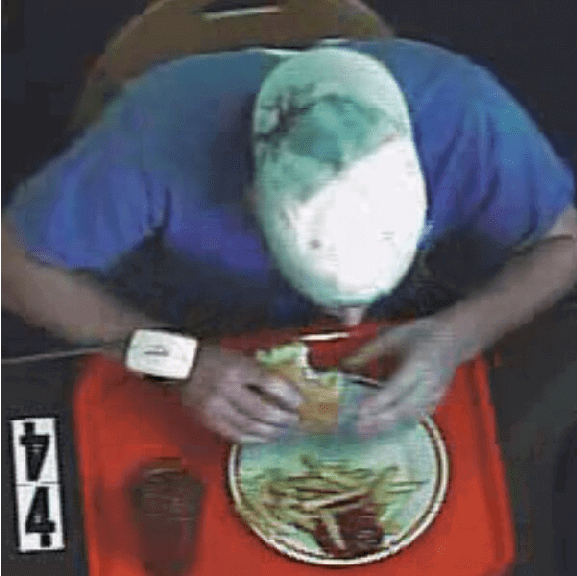} &
\includegraphics[width=0.16\textwidth]{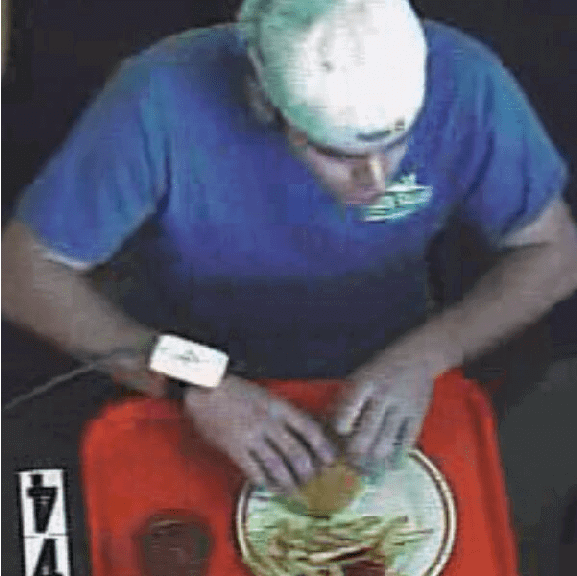} \\
(a) frame=0 &
(a) frame=7 &
(a) frame=14 &
(a) frame=21 &
(a) frame=28 
\end{tabular}
\end{center}
\caption{Example identifying the time index of a bite (frame 14).}
\label{bite-time}
\end{figure*}

The goal of the ground truthing process was to identify the time, food,
hand, utensil and container for each bite.
Because our data set is so large and was collected during natural
(unscripted) eating, the total process took more than 1,000 man-hours of work.
Figure \ref{tool} shows a custom program we built to facilitate the process.
The left panel displays the video while the right panel shows the
synchronized wrist motion tracking data.
Keyboard controls allow for play, pause, rewind and fast forward.
The horizontal scroll bar allows for jumping throughout the recording
and additional keyboard controls allow for jumping to previously labeled
bites.
A human rater annotates a course by watching the video and pausing it at
times when a bite is seen to be taken, using frame-by-frame rewinding
and forwarding to identify the time when food or beverage is placed into
the mouth.
Figure \ref{bite-time} shows an example of a sequence of images
surrounding a bite.
Once the bite time is identified, the rater presses a key to spawn
a pop-up window that allows the user to select from a list of foods
recorded as having been eaten by the participant during the course,
and a list of hand, utensil and container options.
The process of ground truthing a single course took 20-60 minutes.

\begin{figure*}
\begin{center}
\begin{tabular}{c c c c c}
\includegraphics[width=0.15\textwidth]{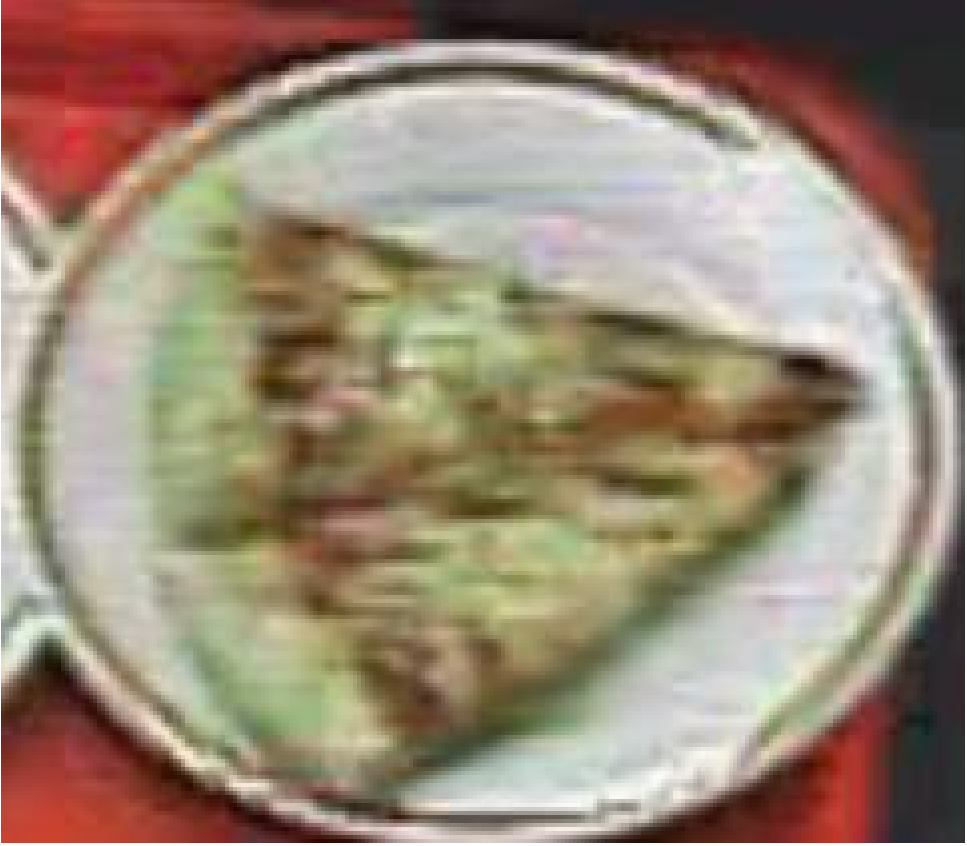} &
\includegraphics[width=0.15\textwidth]{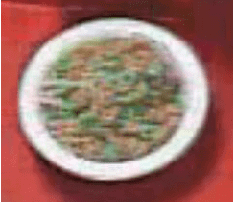} &
\includegraphics[width=0.15\textwidth]{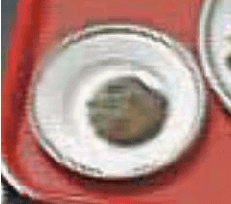} &
\includegraphics[width=0.15\textwidth]{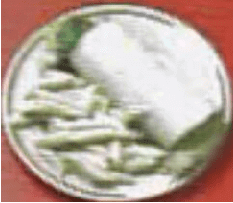} &
\includegraphics[width=0.15\textwidth]{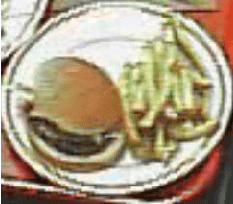}
\end{tabular}
\end{center}
\caption{Examples of foods.  From left to right:
cheese pizza;
cereal Apple Jacks;
chunky chocolate chip cookie;
California chicken wrap, shoestring french fries;
hamburger, shoestring french fries.}
\label{foods-clear}
\end{figure*}

In total, 374 different food and beverage types were chosen by participants.
Food and beverage names were taken from the menus of the cafeteria.
Some foods are given the generic name of the food line from which
they are served due to the heterogeneous
mixture of ingredients that could be custom selected by the participant,
for example from a salad bar.
In cases where a participant mixed 2 or more uniquely chosen foods,
a single name was used that identified the combination.
In cases where a participant ordered a custom version of a food in a
food line, the modifier `custom' was included in the name.
Example food identities include salad bar, shoestring french fries,
Asian vegetables,
pasta tour of Italy, cheese pizza, homestyle chicken sandwich, hamburger,
custom sandwich, garlic breadsticks, fried shrimp and grapefruit.
Example beverage identities include whole milk, coca cola, water,
sweet tea, coffee and apple juice.
Figure \ref{foods-clear} shows some example images of foods.
Foods and beverages were served in four types of containers:  plate,
bowl, glass and mug.
Four different utensils were used:  fork, spoon, chopsticks and hand.
Hand could be identified as left, right or both.

Two human raters independently labeled each course.
A total of 22 raters contributed.

Raters were trained during a 1 hour training session to understand
the process and how to use the program for labeling.  
Quantifying rater agreement is complicated because labeling
is a two step process.
First, each rater had to decide when bites occurred.
Second, they had to quantify food, hand, utensil and container for each bite.
Therefore we developed a two stage approach to determining rater agreement.
For each bite labeled by one rater, a $\pm 1$ sec window was searched for a
corresponding bite from the second rater.
If the food identity, hand, utensil and container all matched, then the
bite was considered matched and the time index was taken as the average
of the time indicated by the two raters.
If a corresponding bite was found within the window but one or more of
the variables did not match, then the bite was reviewed by
a third rater who judged which variable values were correct.
If no corresponding bite was found within the window, the third rater
reviewed the bite to determine if it was missed by one of the raters or
if it was off by more than 1 sec from a bite labeled by the other rater,
in which case the third rater judged the correct time.

\begin{figure*}
\begin{center}
\begin{tabular}{c c c c c}
\includegraphics[width=0.15\textwidth]{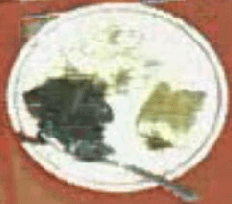} &
\includegraphics[width=0.15\textwidth]{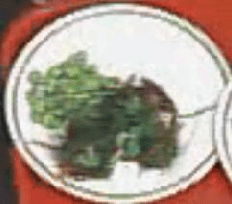} &
\includegraphics[width=0.15\textwidth]{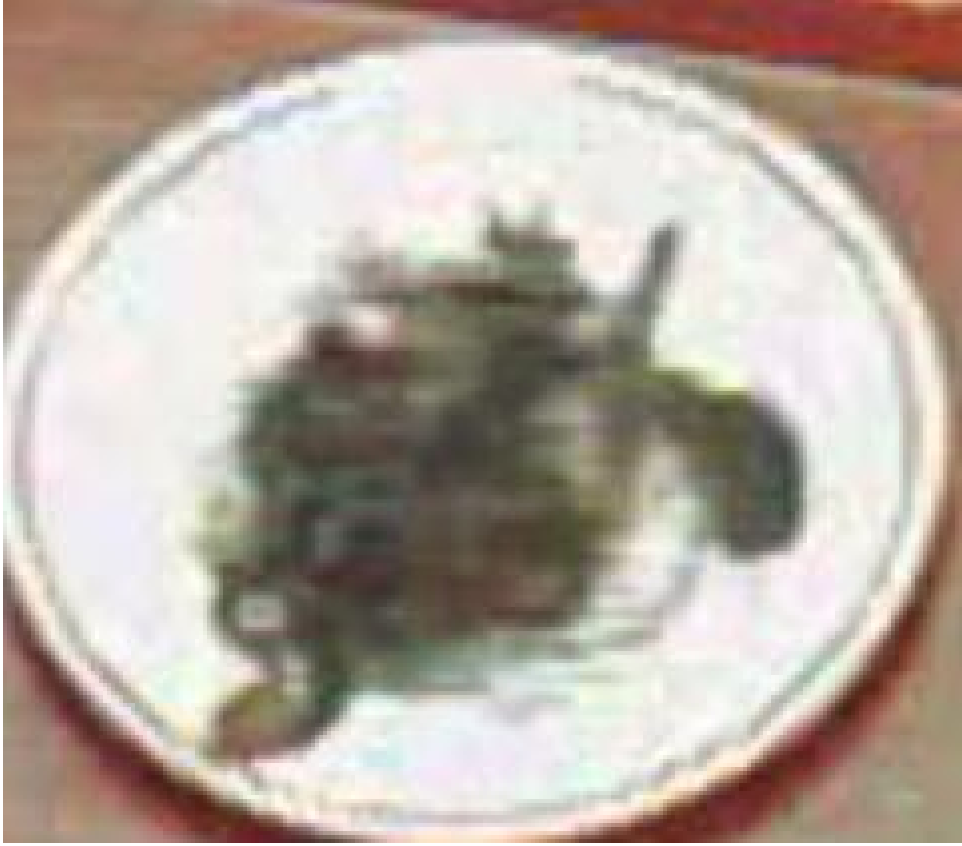} &
\includegraphics[width=0.15\textwidth]{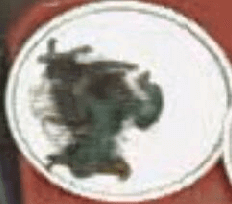} &
\includegraphics[width=0.15\textwidth]{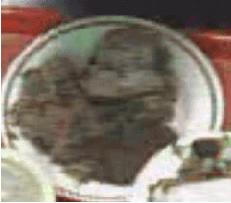}
\end{tabular}
\end{center}
\caption{Examples of foods that are difficult to identify bite by bite.
From left to right:
collard greens, macaroni and cheese, corn bread;
edamame, jasmine rice, stir fry;
char sui braised pork, brown rice, peas and carrots;
pork chop suey with white rice, turkey sliced;
Mexican rice, refried beans, roast pork loin.}
\label{foods-difficult}
\end{figure*}

\begin{figure*}
\begin{center}
\begin{tabular}{c c c c c}
\includegraphics[width=0.16\textwidth]{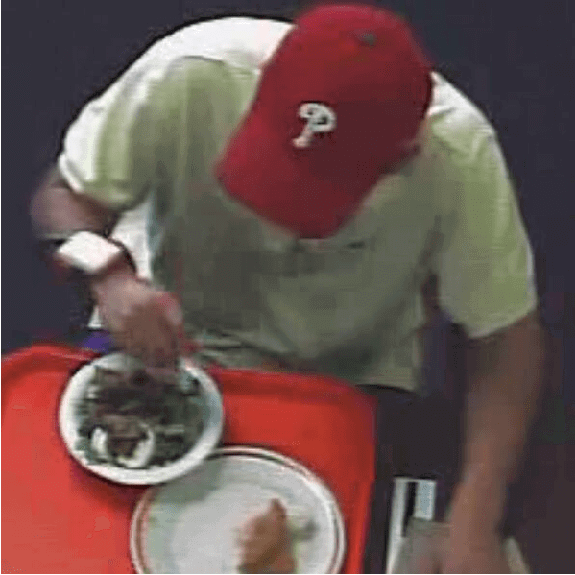} &
\includegraphics[width=0.16\textwidth]{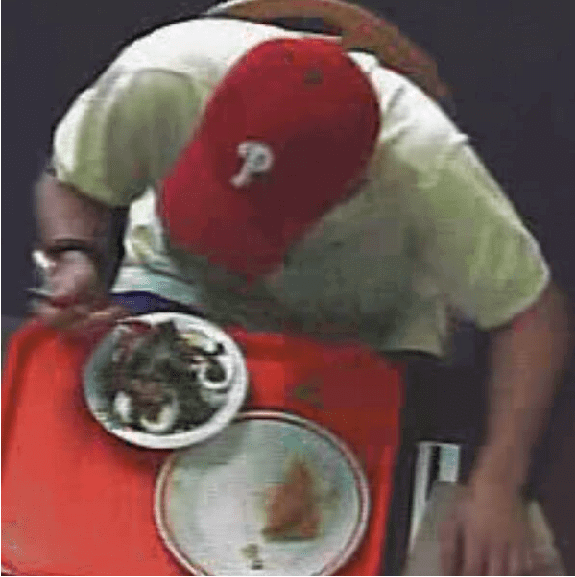} &
\includegraphics[width=0.16\textwidth]{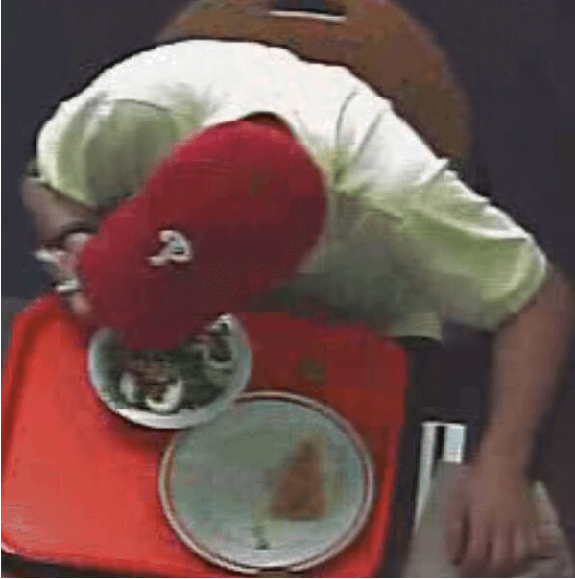} &
\includegraphics[width=0.16\textwidth]{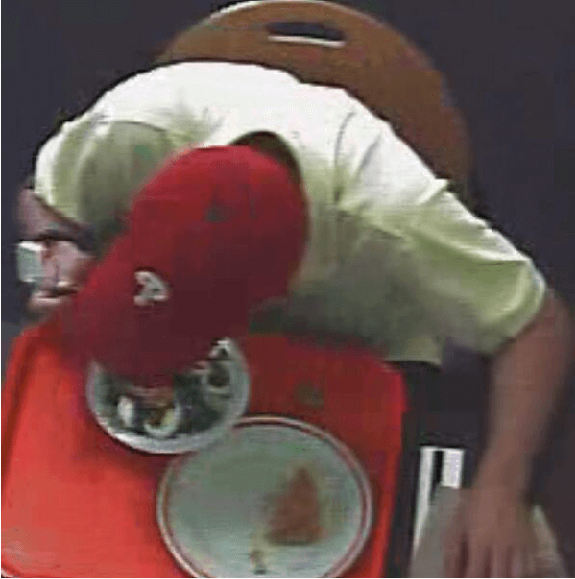} &
\includegraphics[width=0.16\textwidth]{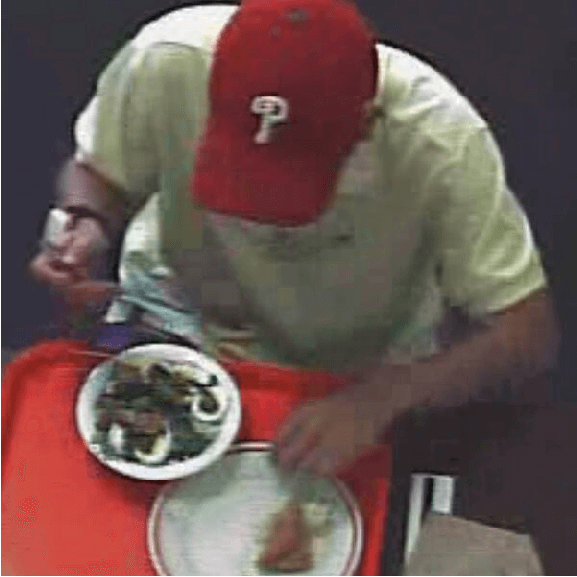} \\
(a) frame=0 &
(a) frame=15 &
(a) frame=30 &
(a) frame=45 &
(a) frame=60 
\end{tabular}
\end{center}
\caption{Example of difficulty identifying the time index of a bite
due to obscuring head motion.}
\label{bite-time-hard}
\end{figure*}

Using this process, rater performance can be evaluated using four metrics:
mistaken identity (food identified incorrectly),
time error (bite labeled more than 1 second from actual time),
missed bite (the rater missed the bite completely)
and data entry error (hand, utensil or container was mislabeled).
Figure \ref{foods-difficult} shows some examples of foods that can be
difficult to identify, for example when 2 or more foods of similar color
and texture are served overlapping each other.
Figure \ref{bite-time-hard} illustrates an example of when the time of
a bite can be difficult to determine due to the head of the participant
obscuring the precise time of food intake.
Data entry errors occurred most commonly when a rater mistakenly
labeled a bowl as a plate or a mug as a glass, either of which would
propagate to all the related bites in the course.
Table \ref{rater-error} summarizes the errors found as judged by the
third rater.

\begin{table}
\begin{center}
\begin{tabular}{|c|c|}
\hline
missed bites & 900 (3.7\%) \\
\hline
time error & 1217 (5\%) \\
\hline
identity error & 714 (3\%) \\
\hline
data entry error & 1059 (4.4\%) \\
\hline
\end{tabular}
\end{center}
\caption{Manual labeling error rates.}
\label{rater-error}
\end{table}

The usefulness of a fourth rater independently labeling each course
and then comparing it to the union judged by the third rater
was explored.  After 71 courses were labeled, the process was stopped.
In those 71 courses the following total errors were found:
17 missed bites, 0 timing errors, 18 identity errors and 8 data entry
errors.  Given the large amount of time needed to independently label
the data and the tiny amount of new errors discovered, it was determined
that the quality of ground truth provided by two human raters and
then judged by a third rater was sufficient.

\subsection{Bite counting algorithm}

The bite counting algorithm described in \cite{Dong12} is briefly
repeated here for background.
The algorithm detects a pattern of wrist roll motion associated with
a bite through the detection of four events.
First, the wrist roll velocity must surpass a positive threshold.
Second, a minimum amount of time must pass.
Third, the velocity must surpass a negative threshold.
Finally, a minimum time must pass between the negative wrist roll for
one bite and the positive wrist roll for the beginning of a next bite.
The minimum times help reduce false positives during other motions.
The algorithm for detecting a bite based on this motion pattern
can be implemented as follows:
\begin{verbatim}
Let EVENT = 0
Loop
  Let Vt = measured roll vel. at time t
  If Vt > T1 and EVENT = 0
    EVENT = 1
    Let s = t
  if Vt < T2 and t-s > T3 and EVENT = 1
    Bite detected
    Let s = t
    EVENT = 2
  if EVENT = 2 and t-s > T4
    EVENT = 0
\end{verbatim}
The variable $EVENT$ iterates through the events just described.
The parameters $T1$ and $T2$ define the threshold for roll detections,
the parameter $T3$ defines the minimum time between positive and
negative rolls, and the parameter $T4$ defines the minimum time between bites.

\subsection{Evaluation metrics}

\begin{figure}
\begin{center}
\includegraphics[width=0.45\textwidth]{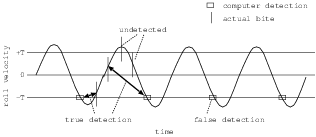}
\end{center}
\caption{Classification of results.}
\label{evaluation}
\end{figure}

The evaluation method follows the procedure previously
established \cite{Dong12}.
Algorithm bite detections are compared to ground
truth manually marked bites.
Figure~\ref{evaluation} illustrates the possible classifications.
For each computer detected bite (small square in the figure),
the interval of time from the previous detection
to the following detection is considered.
The first actual bite taken within this window, that has not yet
been paired with a bite detection, is classified as a true detection (T).
If there are no actual bite detections within that window, then
the bite detection is classified as a false detection (F).
After all bite detections have been classified, any additional actual
bites that remain unpaired to bite detections are classified
as undetected bites (U).
This approach defines an
objective range of time in which an actual bite must have occurred
in order to classify a detected bite as a true positive.
The window extends prior to the actual bite because
it is possible in some cases for the wrist roll motion to
complete just prior to the actual placing of food into the mouth.
Sensitivity (true detection rate) is calculated
as (total Ts)/(total Ts+ total Us).
Because this method does not allow for the definition of a true negative,
specificity (false detection rate) cannot be calculated.
We therefore calculate the positive predictive value as a measure of
performance regarding false positives.
The positive predictive value (PPV) is calculated
as (total Ts)/(total Ts+ total Fs).

\subsection{Parameter Tuning}
\label{parameters}

In the original experiment involving 49 people eating a meal
in a laboratory setting, $T1=T2=10$, $T3=2$ and $T4=8$ were
determined to be optimal \cite{Dong12}.
It was also found that a range of values provided reasonable results.
The present work reports results using these same values but also
reports results using a shorter time for $T4$.
During evaluation is was discovered that people ate faster on average
in the cafeteria experiment than in the previous laboratory experiment.
It was found that setting $T4=6$ produced a more
balanced sensitivity and positive predictive value.
This is further discussed in sections \ref{results}-\ref{discussion}.

\section{Results}
\label{results}

Table \ref{results-demo} lists the sensitivities
found across demographic
variables age, gender and ethnicity.
Sensitivity trended higher as age increased.
Sensitivity for females was 
10\% higher than sensitivity for males.
For ethnicity, sensitivity was highest for
African Americans and lowest for Asians/Pacific Islanders.
Table \ref{results-demo} also reports the average eating rate for
each demographic in seconds per bite (SPB).
SPB trends lower for every demographic as sensitivity trends lower,
suggesting that a faster eating rate results in lower sensitivity.

\begin{table}
\begin{center}
\begin{tabular}{| l | r | r | r | r |}
\hline
demographic & \#partic. & \#bites & \#detected (sensitivity) & SPB \\
\hline 
\hline 
age & &  &  &  \\
\hline 
51-75 & 21 & 1634 & 1404 (86\%)	& 18 \\
\hline
41-50 & 33 & 2790 & 2227 (80\%)	& 17 \\
\hline 
31-40 & 27 & 2531 & 1949 (77\%)	& 15 \\
\hline
24-30 & 76 & 7426 & 5326 (72\%) & 13 \\
\hline
18-23 & 114 & 9707 & 7050 (73\%) & 13 \\
\hline 
\hline 
gender & &  &  &  \\
\hline
female & 140 & 11811 &	9401 (80\%) & 15 \\
\hline
male & 131 & 12277 & 8555 (70\%) & 13 \\
\hline
\hline
ethnicity & & & & \\
\hline
African American & 26 & 1958 & 1583 (81\%) & 18 \\
\hline
Caucasian & 190 & 15990 & 12327 (77\%) & 15 \\
\hline
Hispanic & 11 & 1195 & 877 (73\%) & 13 \\
\hline
Other & 15 & 1635 & 1115 (68\%) & 14 \\
\hline
Asian or Pac. Isl. & 29 & 3310 & 2054 (62\%) & 12 \\
\hline
\end{tabular}
\end{center}
\caption{Sensitivity and seconds per bite (SPB) for age, gender,
and ethnicity.}
\label{results-demo}
\end{table}

\begin{figure*}
\begin{center}
\includegraphics[angle=270]{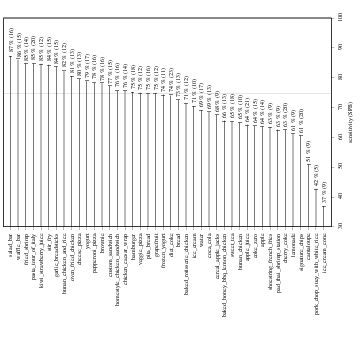}
\caption{Sensitivity and seconds per bite (SPB)
for all foods of which participants consumed
greater than 100 bites.
Frequency (number of occurrences) of bites
for food types in this figure ranged from 110 to 3,986.
Average sensitivity (75\%) highlighted for reference.}
\label{foods_accuracy}
\end{center}
\end{figure*}

Figure \ref{foods_accuracy} plots the sensitivity of the method
for the foods of which more than 100 bites were consumed.
The average sensitivity (75\%) is given for reference.
For most foods the sensitivity trends consistently
in the range of 60-90\%.
For a small number of foods the sensitivity drops precipitously.
For a food like ice cream cone the decrease in sensitivity is likely due
to the natural minimization of wrist roll during consumption
(for fear of having the ice cream fall out of the cone).
Figure \ref{foods_accuracy} also shows the average SPB of each food type.
The correlation between SPB and sensitivity is 0.4 suggesting it
has a mild effect.
To look for other potential causes of variability
we manually observed the motion in the hundreds of hours
of video to try to infer commonalities.
In many cases a bite involves head-towards-plate motion
in combination with hand-towards-mouth motion.
The former seems to be larger when a food is more prone to spillage,
so a participant positions their head over the container to facilitate
delivery of the food to the mouth
(for example, compare figure \ref{bite-time} to figure \ref{bite-time-hard}).
To explore this hypothesis we calculated the amount of motion
of the wrist during a 2 second window centered on every bite and
took the average value for each food type, finding a 0.4 correlation
which again suggests a mild effect.

Table \ref{results-bites} summarizes the accuracies found across other
bite type variables.
Container sensitivity was fairly consistent with the exception of glass
which was 9\% lower than average.
For utensils, chopsticks showed a relatively low detection rate (50\%)
but were also found to be used twice as fast (7 seconds per bite) as
a fork or hand (14-15 seconds per bite).
Handedness showed a small variation in sensitivity, while the use of
both hands as opposed to a single hand reduced sensitivity by 8-9\%.

\begin{table}
\begin{center}
\begin{tabular}{| l | r | r | r |}
\hline
bite variable & \#bites & \#detected (sensitivity) & SPB \\
\hline
\hline
container &  &  &  \\  \hline
bowl     & 3939 & 3091 (79\%) & 15 \\
\hline
mug & 116 & 87 (75\%) & 17 \\
\hline
plate &    16434 &    12389 (74\%)  & 15 \\
\hline
glass &    3599 & 2389 (66\%)  & 19 \\
\hline
\hline
utensil &  &  &    \\
\hline
fork     & 10308 & 8627 (83\%)  & 16 \\
\hline
spoon &    2389 & 1711 (73\%)  & 12  \\
\hline
hand     & 10989 & 7419 (68\%)  & 16  \\
\hline
chopsticks &     400 & 198 (50\%)  & 7  \\
\hline
\hline
hand used &  &  &    \\
\hline
l-handed using left hand & 1363 & 1106 (81\%)  & 15  \\
\hline
r-handed using right hand & 18344 & 14267 (78\%)  & 15  \\
\hline
l-handed using both hands & 162 & 116 (72\%)  & 19 \\
\hline
r-handed using both hands & 1233 & 860 (70\%)  & 16  \\
\hline
\end{tabular}
\end{center}
\caption{Sensitivity and seconds per bite (SPB)
for container, utensils, and hand used.}
\label{results-bites}
\end{table}

Overall, across all 24,088 bites the sensitivity was 75\% with a positive
predictive value of 89\%.
The algorithm parameters were originally determined using
data recorded in a laboratory setting \cite{Dong12} in which the
average eating rate was slower
(n=49, seconds per bite = 19.1 $\pm$ 6.4)
compared to what was observed in the cafeteria setting
(n=271, seconds per bite = 14.7 $\pm$ 5.6).
We therefore experimented with shortening the parameter controlling
the minimum time between detections of bites to 6 seconds.
With this value the algorithm produced 81\% sensitivity with a positive
predictive value of 83\%.

\section{Discussion}
\label{discussion}

The primary goal of this study was to assess the accuracy of the
bite counting method across a wide variety of demographics and food types.
While minor variations occurred across most variables, the method
showed robustness to this challenging data set.
The original laboratory test found 81\% sensitivity with 86\% positive
predictive value \cite{Dong12}.
After tuning the algorithm to the faster eating pace observed in the
cafeteria, the same sensitivity was achieved with only a 3\% decrease
in positive predictive value.
This experiment provides the most comprehensive evidence to date that
the method is reliable during normal unscripted eating.

The experiment identified two areas where the algorithm could be improved.
First, variations in eating pace affect the sensitivity.
The bite detection algorithm includes a parameter (T4)
that defines the minimum time between bites.
It is intended to reduce false positives that may be caused
by non-eating wrist motions.
In our previous experiment in a
laboratory (49 people), we found that tuning T4 to 8 seconds provided the
best average results \cite{Dong12}. In the cafeteria experiment
reported in this paper (271 people), we found that tuning T4 to 6 seconds
provided the best average results.
We also found that there were some differences in average eating
rate across demographic variables (age, gender, ethnicity)
that trended with bite detection sensitivity.
In future work we intend to use those demographic variables to try
to automatically adjust T4.
We also intend to try to detect eating rate from the wrist motion
tracking signals to automatically adjust to the individual.  
This would be similar to how a pedometer learns the stride
duration of a person while running or walking and adjusts its step
detection parameters accordingly.
Second, variations in the amount of wrist motion versus the amount
of head-towards-plate motion affect the sensitivity.
Two parameters of the algorithm are designed to detect the typical
amount of motion.  Again it may be possible to adjust these parameters
in real-time to learn the typical amount of wrist motion of a person
during a meal.
This work provides the data set necessary to explore these ideas.

One limitation of the bite counting algorithm is that it requires
a user to turn the method on/off at the beginning/end of a meal.
However, in a previous study we analyzed data from 77 participants
consuming 2,975 meals over a 2 week period \cite{Scisco14}.
This demonstrated good compliance with remembering to use the device.
Another potential limitation of the bite counting algorithm is
its susceptibility to false positives caused by wrist motions unrelated
to eating.
However, in this experiment we did not script the eating activity
or restrict the types of motions of the participants.
People were instructed to eat as naturally as possible and thus
the amount of non-eating wrist motions can be expected to be typical.
In our previously published laboratory experiment,
we manually reviewed the videos and counted non-eating
wrist motions such as those caused by using a napkin, phone, or engaging
in conversation, and found that they occurred between 67\% of bites.
Collectively our experiments demonstrate robustness to typical non-eating
wrist motions during normal eating.

A strength of the experiment reported in this paper is that the
eating recorded took place in an environment that was as natural
as possible, and eating behaviors were completely unscripted and unrestricted.
A weakness of this approach is that it requires a tremendous effort
in labeling ground truth.  In total over 1,000 man hours were
invested in reviewing the videos and labeling the bites.
We recruited 22 reviewers because of the large effort needed to
complete the ground truthing process.
Studies have shown that participants change their eating behavior
in clinical settings \cite{deCastro00,Petty13}.
As this method is intended to be used in free-living scenarios,
a naturalistic evaluation of its accuracy is important.
However, although we tried to make the cafeteria setting as natural
as possible, it is still possible that behaviors in free-living
environments could affect the accuracy of the method in ways that
could not be captured with this study (e.g. grazing, other types
of distraction).
Future studies should examine the algorithm's accuracy in these types of situations.

\section{Acknowledgments}

We gratefully acknowledge the support of the NIH via grants
1R41DK091141-A1 and 2R42DK091141-02.
We also wish to thank the 22 volunteers who manually labeled the
bite database for their hundreds of hours of work.

\vfill
\end{document}